\documentclass[11pt,a4,twoside]{article}

\usepackage{./styling/style-main}
\graphicspath{{plots/}}
\usepackage{framed}
\usepackage{amsmath}
\usepackage{slashed}

\makeatletter
\renewcommand\paragraph{\@startsection{paragraph}{4}{\z@}%
	{3.25ex \@plus1ex \@minus.2ex}%
	{-1em}%
	{\normalfont\normalsize\bfseries}}
\makeatother

\newcommand{\bracket}[1]{\left\langle #1 \right\rangle}

\usepackage{tikz-feynman}
\usepackage{rotating}
\usepackage{transparent}

	\title{Two-loop hard thermal loops for any model}

\author{Andreas~Ekstedt\thanks{andreas.ekstedt@desy.de}\textsuperscript{~~,\,a,\,b\,c}}
\affil{a:~II. Institute of Theoretical Physics, Universität Hamburg, D-22761, Hamburg, Germany}
\affil{b:~Deutsches Elektronen-Synchrotron DESY, Notkestr. 85, 22607 Hamburg, Germany}
\affil{c:~Department of Physics and Astronomy, Uppsala University, P.O. Box 256, SE-751 05 Uppsala, Sweden}

\date{\today}
\begin{document}
	\begin{flushright}
		\footnotesize
		{\large DESY-23-016} \\
	\end{flushright}
{\let\newpage\relax\maketitle}
	\thispagestyle{plain}
	\begin{abstract}
	 Hard thermal loops describe how soft gauge fields are screened and damped in hot plasmas. As such they are used to calculate transport coefficients, Sphaleron rates, equations of state, and particle production. However, most calculations are done using one-loop self-energies. And two-loop contributions can be large. To that end this paper provides vector two-loop self-energies for generic models: Any scalar, fermion, or vector representation; and all possible renormalizable terms. Several examples are given to showcase the results. Two-loop results for higher-point functions are also given.
	\end{abstract}

\section{Introduction}
Be it phase transitions~\cite{Arnold:1992rz,Kajantie:1995kf,Moore:2000jw}; Baryon violation~\cite{Moore:1998swa,Arnold:1996dy}; photon emission from heavy-ion collisions~\cite{Arnold:2001ba,Arnold:2001ms}; or axion production~\cite{Salvio:2013iaa,RevModPhys.53.43,Borsanyi:2016ksw}; thermal field theory is indispensable all the same. Whilst the picture is quite complicated for generic systems, the physics is considerably simpler if we look at length-scales of the order $L\gg T^{-1}$. For in that case high-energy modes with $E\sim T$ behave as quasiparticles~\cite{Arnold:1997gh,Blaizot:1993zk}. And much intuition from plasma physics directly carries over. So as charged particles move in, say an electric field, they redistribute themselves to screen the field. Likewise, free charges resist a changing magnetic field in accordance with Lenz's law. In both cases the field is screened by high-energy modes.

The best way to incorporate this screening depends on the situation. In equilibrium, for example, the scalar potential  ($A^0$) effectively obtains a thermal Debye mass~\cite{Arnold:1992rz}. Since there is no time dependence, it is useful to describe such systems with a three-dimensional field theory~\cite{Kajantie:1995dw,Farakos:1994kx}. A different effective description,  known as hard thermal loops, can be used when fields vary slowly in time.

These hard thermal loops are particularly important when the system is pushed from equilibrium. This is because deviations from equilibrium are driven back by scattering processes; and the characteristic momentum transfer, and thus the cross-section, is set by the screening length. As such hard thermal loops are key for calculating transport coefficients~\cite{Jeon:1994if,Arnold:2006fz,Arnold:2003zc,Arnold:2000dr,Jeon:1995zm}, particle production~\cite{Ghiglieri:2020mhm,Arnold:2002ja,Ghiglieri:2022rfp}, and colour conductivity~\cite{Bodeker:1999ey,Arnold:1998cy,Arnold:1999uy}.

Though hard thermal loops are important, little is known about them beyond leading order. Existing studies are limited to quantum electrodynamics at high temperatures~\cite{Carignano:2017ovz,Carignano:2019ofj} and at finite chemical potential~\cite{Gorda:2022fci}. Reason being that direct evaluations are hampered by an increased complexity at two loops. Nevertheless, in this paper we use kinetic theory to simplify the calculations. Our method of choice is rather compact and admits neat expressions for generic model\te including non-abelian theories.

The first section of the paper describes the calculation; section \ref{sec:Generic} provides results for general models; section \ref{sec:HigherPoint} provides higher-point correlators; and additional details are given in the appendices. The results are also given in the accompanying {\tt HTLGen.m} file.

\section{The real-time formalism}
Throughout this article we use the mostly-plus metric: $P^2=-(p^0)^2+\vec{p}^2$, and all four-vectors are denoted by capitalized letters, while spatial vectors are denoted by lowercase ones. To save ink we also use the notation $p^2 \equiv \vec{p}^2$.

Because we are interested in real-time dynamics we have to double the field content~\cite{Niemi:1983nf,Niemi:1983ea}: Here we follow~\cite{CHOU19851,Caron-Huot:2007cma,Arnold:2001ba} and use retarded and advanced fields, otherwise known as the r/a basis. In this basis there are three propagators for each field. For a free theory these are\footnote{See \cite{Caron-Huot:2007cma,Caron-Huot:2007zhp} for a clear diagrammatic representation of Feynman rules in this basis.}
\begin{align}
&\Delta^{rr}_{B/F}(P)= 2\pi \delta(P^2)\left\{ \theta(p^0)N^{+}_{B/F}(p^0,\vec{p})+\theta(-p^0) N^{-}_{B/F}(-p^0,-\vec{p})\right\}, 
\\& \Delta^R(P)=\frac{-i}{P^2-i \eta p^0}, \quad \Delta^A(P)=\frac{-i}{P^2+i \eta p^0},\quad N_{B/F}(p^0,\vec{p})=\frac{1}{2}\pm n_{B/F}(p^0).
\end{align}
To condense the notation we denote $rr$ propagators by
\begin{align}
	&\Delta_X(P)= 2\pi \delta(P^2)\left\{ \theta(p^0) N_X(p^0,\vec{p})+\theta(-p^0) \overline{N}_X(-p^0,-\vec{p})\right\}, 
\end{align}
where $X=V,F,S$ depending on the particle. In this case the $rr$ propagator for vectors, fermions, and scalars is
\begin{align}
	D^{rr}_{\mu \nu}(P)=g_{\mu \nu} \Delta_V(P), \quad S_F^{rr}=-\slashed{P} \Delta_F(P), \quad D_S^{rr}(P)=\Delta_S(P),
\end{align}
where we have used Feynman gauge.

To handle divergences we use dimensional regularization. This means that our integration measures are
\begin{align}
\int_P \equiv  \left(\frac{\mu^2 e^{\gamma}}{4\pi}\right)^\epsilon\int \frac{d^D P}{(2\pi)^D}, \quad \int_p \equiv  \left(\frac{\mu^2 e^{\gamma}}{4\pi}\right)^\epsilon\int \frac{d^d p}{(2\pi)^d},
\end{align}
where $D=4-2\epsilon$ and $d=3-2\epsilon$.

\subsection{Hard thermal loops from transport equations}
As of yet, two-loop hard thermal loops are only known for quantum electrodynamics~\cite{Carignano:2019ofj,Carignano:2017ovz,Gorda:2022fci}. These calculations are quite involved and have so far been done using Feynman diagrams\footnote{See \cite{Jackson:2019mop} for results with general external momenta.}. To make our calculations tractable, we instead use transport equations. This method is well-known, and is a clean way to derive hard thermal loops at leading order~\cite{Blaizot:1993zk,Blaizot:1993be,Litim:2001db,Blaizot:2001nr}. Here we extend the method to the next order. Essentially we use that fields with typical momenta $p\sim T$ can be treated as quasiparticles. For example, we can describe electrons with the Vlasov equation:
\begin{align}
 \dot{N}_F^{\pm}+\vec{v}\cdot \vec{\nabla}N_F^{\pm}\pm e \left(\vec{E}+\vec{v}\times \vec{B}\right)\cdot \vec{\nabla}^p N_F^{\pm}=0.
\end{align}
If we now assume that the electrons are driven slightly away from equilibrium by the electric field, we can expand the electron distribution as 
\begin{align}\label{eq:Vlasov}
&N_F^{\pm}=\frac{1}{2}-n_F-\delta n_F^{\pm}, \quad v\cdot \partial \delta n_F^{\pm}(\vec{p},x)=\mp e \vec{v}\cdot \vec{E} \frac{d}{d p} n_F(p),
\\& v^{\mu}=(1,\vec{v}), \quad \vec{v}\equiv \frac{\vec{p}}{p^0}, \quad n_F(p)=\left(e^{p/T}+1\right)^{-1}.
\end{align}

The photon self-energy then follows from the electron current~\cite{Blaizot:1993zk,Blaizot:1993be}:
\begin{align}
\partial_\mu F^{\nu \mu}=e \bracket{\bar{\Psi} \gamma^\nu \Psi}\sim e\int_p v^\mu (N^{+}_F-N^{-}_F),
\end{align}
where $\bracket{.,.}$ denotes the average over hard modes with characteristic momenta $p \sim T$. 
That is
\begin{align}
&\partial_\mu F^{\nu \mu}=-e \int \frac{d^4p}{(2\pi)^4}\text{Tr}\left[\slashed{p}\gamma^\nu\right] \Delta_F(p)=2 e \int \frac{d^3p}{(2\pi)^3} v^{\nu} \left[N^{+}(p,x)-N^{-}(p,x)\right]
\\&=4 e^2 \int \frac{d^3p}{(2\pi)^3}\frac{ v^{\nu} \vec{v}\cdot\vec{E}(x)}{v\cdot K} n'_F(p)=-\frac{e^2T^2}{3} \int \frac{d\Omega_v}{4\pi} \frac{v^{\nu} \vec{v}\cdot\vec{E}(x) }{v\cdot K}\equiv \Pi^{\nu \mu}A_\mu,
\end{align}
where $\vec{E}=-\vec{\nabla}A^0-\dot{\vec{A}}$. 

Note that the kinetic approach works because quantum fields with $p \sim e T$ behave classically at high temperatures. In generic situations we have no right to expect classical equations of motion. We should also remember that scattering processes become important at time scales of order $t\sim (e^4T)^{-1}$\cite{Bodeker:1999ey,Arnold:1997gh,Arnold:1998cy,Arnold:1999uy}, and that our results only hold for soft fields: $\dot{A}\sim \vec{\nabla} A \sim (e T) A$.

\subsection{Using kinetic theory beyond leading order}
There are two ways that we can go about applying the kinetic approach at two loops. First, we can include resummed self-energies directly in the transport equations and use this to derive effective particle distributions~\cite{Blaizot:2001nr}. While possible, this approach involves evaluating self-energies at finite external momentum. Instead we elect to only use leading-order transport equations\te two-loop results are then obtained by calculating corrections to the fermion current $ \bracket{\bar{\Psi} \gamma^\nu \Psi}$. At first glance it seems like we are back to brute-force evaluating diagrams. Be that as it may, working with currents is considerably easier than calculating self energies. And as we shall see, the results for different kinds of particles involve the same compact expressions.

\subsection{Two-loop hard thermal loops}

Let us demonstrate our approach for quantum electrodynamics. The two-loop contribution to the electron current is shown in figure \ref{fig:FermionVector}:
\begin{align}\label{eq:QEDSelfEnergy}
 \bracket{\bar{\Psi} \gamma^\mu \Psi}_\text{2-loop}=e^2\int_{PQ} F^{\mu}&\left\{ \Delta_F(P)\Delta_V(Q) \left[ \Delta^R(P)\Delta^R(P+Q)+\Delta^A(P)\Delta^A(P+Q)\right]\right.\nonumber
\\&\left. +\Delta_F(P)\Delta_F(P+Q)\left[\Delta^R(P)\Delta^A(Q)+\Delta^A(P)\Delta^R(Q)\right] \right. \nonumber
\\& \left.  +\Delta_F(P+Q) \Delta_V(Q)\left[\Delta^R(P)\Delta^A(P)\right]\right\},
\end{align}
where
$F^\mu= \text{Tr} \slashed{P}\gamma^{\mu}\slashed{P}\gamma^\alpha \left(\slashed{P}+\slashed{Q}\right)\gamma_\alpha=-4(D-2)\left[(P+Q)^2 p^\mu-P^2 q^\mu -Q^2 p^\mu\right]$.

To evaluate the integrals we have to define
\begin{align}
	\delta(p^2) \Delta^{R/A}(P).
\end{align}
This expression contains terms with two delta functions\te these must be regulated. To do so we use the original approach~\cite{Niemi:1983ea}:
\begin{align}\label{eq:DoubleDelta}
	& \pi \delta(P^2)\Delta^{R/A}(P)=\frac{-i}{P^2\mp i \eta p^0}\frac{\eta}{P^4+\eta^2}=\pm p^0 \left[\frac{\eta}{P^4+\eta^2}\right]^2 - i\frac{P^2 \eta}{(P^4+\eta^2)^2}
	\\&=\pm p^0 \left[\pi \delta(P^2)\right]^2- \frac{i}{2}\frac{\partial}{\partial p_0^2}\left[\pi \delta(P^2)\right].
\end{align}
For a given topology all $\left[\pi \delta(P^2)\right]^2$ terms cancel, while the remaining pieces can be handled by integration-by-parts. 

As an example, consider
\begin{align}
\int_{PQ}F^{\mu}\Delta_F(P)\Delta_V(Q) \left[ \Delta^R(P)\Delta^R(P+Q)+\Delta^A(P)\Delta^A(P+Q)\right].
\end{align}
After using equation \eqref{eq:DoubleDelta} we find
\begin{align}
&\pi \Delta_F(P^2) \left[ \Delta^R(P)\Delta^R(P+Q)+\Delta^A(P)\Delta^A(P+Q)\right]
\\&=p^0\left[\pi \delta(P^2)\right]^2\left[\Delta^R(P+Q)-\Delta^A(P+Q)\right]-\frac{i}{2}\frac{\partial}{\partial p_0^2}\left[\pi \delta(P^2)\right]\left[\Delta^R(P+Q)+\Delta^A(P+Q)\right]. \nonumber
\end{align}
The first term vanishes, so we are left with the second term. Now, for the $P^2 q^\mu$ term to contribute, the $\frac{\partial}{\partial p_0^2}$ derivative must hit $P^2$. So this term is proportional to
\begin{align}
\int_{PQ} q^{\mu}  \Delta_F(P)\Delta_V(Q) \left[\frac{1}{(P+Q)^2}\right].
\end{align}
Naively we expect a collinear ($\vec{p}\parallel \vec{q}$) divergence from the angular integration, but these cancel once we sum all contributions.

The $p^\mu (P+Q)^2$ factor results in a term proportional to
\begin{align}
\int_{PQ} \frac{N_V(q)}{q p}\left\{\left[\partial^p_0 N_F-\partial^p_0\overline{N}_F\right]v^\mu_p-\frac{v^\mu_p-n^{\mu}}{p}(N_F-\overline{N}_F)\right\}, \quad n^{\mu}=\left(1,\vec{0}\right).
\end{align}
Finally, the $Q^2 p^\mu$ term does not contribute as $\Delta_V(Q)$ sets $Q^2=0$.

The remaining terms in $ \bracket{\bar{\Psi} \gamma^\mu \Psi}_\text{2-loop}$ are obtained in the same way. After performing the integrals and using the formulas in appendix \ref{app:DerivDistri}, we find
\begin{align}
	\Pi^{\mu \nu}_\text{NLO}(K)=-\frac{e^4 T^2}{8 \pi^2}\int \frac{d\Omega_v}{4\pi}\left\{v^\mu v^\nu \left[\frac{(k^0)^2}{(v\cdot K)^2}-\frac{2 k^0}{v\cdot K}\right]+\left[v^\mu n^\nu+n^\mu v^\nu\right]\frac{ k^0}{v\cdot K}-n^\mu n^\nu\right\},
\end{align}
which can be compared with the leading-order self-energy
\begin{align}
\Pi^{\mu \nu}_\text{LO}(K)=-\frac{ e^2 T^2}{3} \int \frac{d\Omega_v}{4\pi}\left[n^\mu n^\nu+v^\mu v^\nu \frac{k_0}{v\cdot K} \right].
\end{align}
This result is in agreement with previous calculations ~\cite{Gorda:2022fci,Carignano:2019ofj}. For completeness we have to add power-corrections. This is done in section \ref{sec:PowerCorrection}.

\subsection{Procedure for general diagrams}
Irrespective of the diagram or particle, the only terms that contribute are of the form
\begin{align}
\int_{PQ} (a p^\mu+b q^\mu)(P+Q)^2\Delta_X(P)\Delta_Y(Q) \left[ \Delta^R(P)\Delta^R(P+Q)+\Delta^A(P)\Delta^A(P+Q)\right].
\end{align}
The piece going with $p^\mu$ give terms proportional to
\begin{align}
\int_{PQ}\frac{N_Y(q)}{q p}\left\{\left[\partial^p_0 N_X(p)-\partial^p_0\overline{N}_X(p)\right]v_p^\mu-\frac{v_p^\mu-n^{\mu}}{p}(N_X(p)-\overline{N}_X(p))\right\},
\end{align}
and the term multiplying $q^\mu$ give terms of the form
\begin{align}\label{eq:SecondStruct}
	\int_{PQ}v^\mu_q\left(N_Y(q)-\overline{N}_Y(q)\right)\frac{1}{p^2}\left\{\partial^p_0\left[N_X(p)+\overline{N}_X(p)\right]-\frac{1}{p}\left(N_X(p)+\overline{N}_X(p)\right)\right\}.
\end{align}
In our example we only had the first type, but the second type of terms appears in non-abelian theories. Physically the first Lorentz structure corresponds to deviations from the ballistic approximation: 
\begin{align}
	v^\mu_p= \frac{p^\mu}{p^0}\rightarrow v_p^\mu-\frac{m^2}{2 p^2}\left[v_p^\mu-n^\mu\right]+\ldots,
\end{align}
where $m^2\sim \int p^{-1} n_{B/F}(p)$ represents hard charges obtaining a thermal mass.

The second structure, on the other hand, represents a renormalization of the hard distributions themselves. Connected with this the momentum integral in equation \eqref{eq:SecondStruct} contain divergences\footnote{These cancel once counter-term insertions are added.}.

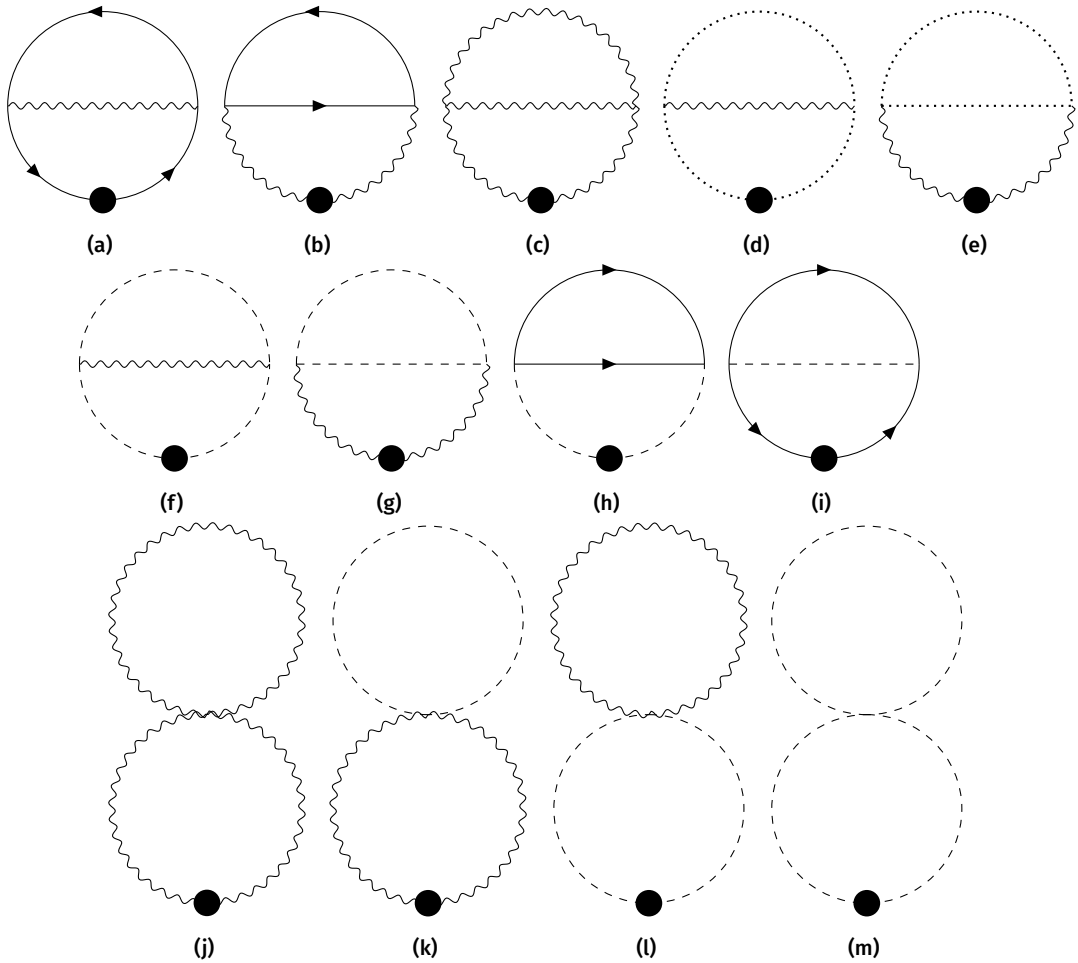
\begin{figure}[h!]
	\centering
	\subfloat[]{
		\begin{tikzpicture}[scale=2.50]
			\begin{feynman} [inline=a]
				\diagram [horizontal=a to b] {			
				};
				\coordinate (b) at (0,0); 
				\coordinate (a) at (1,0);
				\coordinate (c) at (0.5,-0.5);
				\fill (c) circle (2pt);	
				\draw [photon](b) -- (a);
				\draw [anti fermion] (b) arc [start angle=180, end angle=0, radius=0.5cm];
				\draw [fermion] (b) arc [start angle=-180, end angle=-90, radius=0.5cm];
				\draw [fermion] (c) arc [start angle=-90, end angle=0, radius=0.5cm];			
			\end{feynman}
		\end{tikzpicture}
		\label{fig:FermionVector}
	}
	\subfloat[]{
		\begin{tikzpicture}[scale=2.50]
			\begin{feynman} [inline=a]
				\diagram [horizontal=a to b] {			
				};
				\coordinate (b) at (0,0); 
				\coordinate (a) at (1,0);
				\coordinate (c) at (0.5,-0.5);
				\fill (c) circle (2pt);	
				\draw [fermion](b) -- (a);
				\draw [anti fermion] (b) arc [start angle=180, end angle=0, radius=0.5cm];
				\draw [photon] (b) arc [start angle=-180, end angle=-90, radius=0.5cm];
				\draw [photon] (c) arc [start angle=-90, end angle=0, radius=0.5cm];			
			\end{feynman}
		\end{tikzpicture}
		\label{fig:VectorFermion}
	}
	\subfloat[]{
		\begin{tikzpicture}[scale=2.50]
			\begin{feynman} [inline=a]
				\diagram [horizontal=a to b] {			
				};
				\coordinate (b) at (0,0); 
				\coordinate (a) at (1,0);
				\coordinate (c) at (0.5,-0.5);
				\fill (c) circle (2pt);	
				\draw [photon](b) -- (a);
				\draw [photon] (b) arc [start angle=180, end angle=0, radius=0.5cm];
				\draw [photon] (b) arc [start angle=-180, end angle=-90, radius=0.5cm];
				\draw [photon] (c) arc [start angle=-90, end angle=0, radius=0.5cm];			
			\end{feynman}
		\end{tikzpicture}
		\label{fig:VectorVector}
	}
	\subfloat[]{
		\begin{tikzpicture}[scale=2.50]
			\begin{feynman} [inline=a]
				\diagram [horizontal=a to b] {			
				};
				\coordinate (b) at (0,0); 
				\coordinate (a) at (1,0);
				\coordinate (c) at (0.5,-0.5);
				\fill (c) circle (2pt);	
				\draw [photon](b) -- (a);
				\draw [ghost] (b) arc [start angle=180, end angle=0, radius=0.5cm];
				\draw [ghost] (b) arc [start angle=-180, end angle=-90, radius=0.5cm];
				\draw [ghost] (c) arc [start angle=-90, end angle=0, radius=0.5cm];			
			\end{feynman}
		\end{tikzpicture}
		\label{fig:GhostVector}
	}
	\subfloat[]{
		\begin{tikzpicture}[scale=2.50]
			\begin{feynman} [inline=a]
				\diagram [horizontal=a to b] {			
				};
				\coordinate (b) at (0,0); 
				\coordinate (a) at (1,0);
				\coordinate (c) at (0.5,-0.5);
				\fill (c) circle (2pt);	
				\draw [ghost](b) -- (a);
				\draw [ghost] (b) arc [start angle=180, end angle=0, radius=0.5cm];
				\draw [photon] (b) arc [start angle=-180, end angle=-90, radius=0.5cm];
				\draw [photon] (c) arc [start angle=-90, end angle=0, radius=0.5cm];			
			\end{feynman}
		\end{tikzpicture}
		\label{fig:VectorGhost}
	}
	\newline
	\subfloat[]{
		\begin{tikzpicture}[scale=2.50]
			\begin{feynman} [inline=a]
				\diagram [horizontal=a to b] {			
				};
				\coordinate (b) at (0,0); 
				\coordinate (a) at (1,0);
				\coordinate (c) at (0.5,-0.5);
				\fill (c) circle (2pt);	
				\draw [photon](b) -- (a);
				\draw [scalar] (b) arc [start angle=180, end angle=0, radius=0.5cm];
				\draw [scalar] (b) arc [start angle=-180, end angle=-90, radius=0.5cm];
				\draw [scalar] (c) arc [start angle=-90, end angle=0, radius=0.5cm];			
			\end{feynman}
		\end{tikzpicture}
		\label{fig:ScalarVector}
	}
	\subfloat[]{
		\begin{tikzpicture}[scale=2.50]
			\begin{feynman} [inline=a]
				\diagram [horizontal=a to b] {			
				};
				\coordinate (b) at (0,0); 
				\coordinate (a) at (1,0);
				\coordinate (c) at (0.5,-0.5);
				\fill (c) circle (2pt);	
				\draw [scalar](b) -- (a);
				\draw [scalar] (b) arc [start angle=180, end angle=0, radius=0.5cm];
				\draw [photon] (b) arc [start angle=-180, end angle=-90, radius=0.5cm];
				\draw [photon] (c) arc [start angle=-90, end angle=0, radius=0.5cm];			
			\end{feynman}
		\end{tikzpicture}
		\label{fig:VectorScalar}
	}
	\subfloat[]{
		\begin{tikzpicture}[scale=2.50]
			\begin{feynman} [inline=a]
				\diagram [horizontal=a to b] {			
				};
				\coordinate (b) at (0,0); 
				\coordinate (a) at (1,0);
				\coordinate (c) at (0.5,-0.5);
				\fill (c) circle (2pt);	
				\draw [fermion](b) -- (a);
				\draw [fermion] (b) arc [start angle=180, end angle=0, radius=0.5cm];
				\draw [scalar] (b) arc [start angle=-180, end angle=-90, radius=0.5cm];
				\draw [scalar] (c) arc [start angle=-90, end angle=0, radius=0.5cm];			
			\end{feynman}
		\end{tikzpicture}
		\label{fig:ScalarFermion}
	}
	\subfloat[]{
		\begin{tikzpicture}[scale=2.50]
			\begin{feynman} [inline=a]
				\diagram [horizontal=a to b] {			
				};
				\coordinate (b) at (0,0); 
				\coordinate (a) at (1,0);
				\coordinate (c) at (0.5,-0.5);
				\fill (c) circle (2pt);	
				\draw [scalar](b) -- (a);
				\draw [fermion] (b) arc [start angle=180, end angle=0, radius=0.5cm];
				\draw [fermion] (b) arc [start angle=-180, end angle=-90, radius=0.5cm];
				\draw [fermion] (c) arc [start angle=-90, end angle=0, radius=0.5cm];			
			\end{feynman}
		\end{tikzpicture}
		\label{fig:FermionScalar}
	}
	\newline
	\subfloat[]{
		\begin{turn}{90}
			\begin{tikzpicture}[scale=2.50]
				\begin{feynman} [inline=a]
					\diagram [horizontal=a to b] {			
					};
					\coordinate (c) at (-1,0);
					\coordinate (b);
					\fill (c) circle (2pt);	
					\draw [photon] (b) arc [start angle=180, end angle=-180, radius=0.5cm];
					\draw [photon] (b) arc [start angle=0, end angle=-360, radius=0.5cm];		
				\end{feynman}
			\end{tikzpicture}
		\end{turn}
		\label{fig:VectorVectorBubble}
	}
	\subfloat[]{
		\begin{turn}{90}
			\begin{tikzpicture}[scale=2.50]
				\begin{feynman} [inline=a]
					\diagram [horizontal=a to b] {			
					};
					\coordinate (c) at (-1,0);
					\coordinate (b);
					\fill (c) circle (2pt);	
					\draw [scalar] (b) arc [start angle=180, end angle=-180, radius=0.5cm];
					\draw [photon] (b) arc [start angle=0, end angle=-360, radius=0.5cm];		
				\end{feynman}
			\end{tikzpicture}
		\end{turn}
		\label{fig:VectorScalarBubble}
	}
	\subfloat[]{
		\begin{turn}{90}
			\begin{tikzpicture}[scale=2.50]
				\begin{feynman} [inline=a]
					\diagram [horizontal=a to b] {			
					};
					\coordinate (c) at (-1,0);
					\coordinate (b);
					\fill (c) circle (2pt);	
					\draw [photon] (b) arc [start angle=180, end angle=-180, radius=0.5cm];
					\draw [dashed] (b) arc [start angle=0, end angle=-360, radius=0.5cm];		
				\end{feynman}
			\end{tikzpicture}
		\end{turn}
		\label{fig:ScalarVectorBubble}
	}
	\subfloat[]{
		\begin{turn}{90}
			\begin{tikzpicture}[scale=2.50]
				\begin{feynman} [inline=a]
					\diagram [horizontal=a to b] {			
					};
					\coordinate (c) at (-1,0);
					\coordinate (b);
					\fill (c) circle (2pt);	
					\draw [scalar] (b) arc [start angle=180, end angle=-180, radius=0.5cm];
					\draw [dashed] (b) arc [start angle=0, end angle=-360, radius=0.5cm];		
				\end{feynman}
			\end{tikzpicture}
		\end{turn}
		\label{fig:ScalarScalarBubble}
	}
	\caption{Figures \protect\subref{fig:FermionVector} and \protect\subref{fig:FermionScalar} represent corrections to the fermion current; figures \protect\subref{fig:VectorFermion},  \protect\subref{fig:VectorVector}, \protect\subref{fig:VectorGhost}, \protect\subref{fig:VectorScalarBubble}, \protect\subref{fig:VectorVectorBubble},  and \protect\subref{fig:VectorGhost} represent corrections to the vector current; figure \protect\subref{fig:GhostVector} represents corrections to the ghost current; figures  \protect\subref{fig:ScalarVector}, \protect\subref{fig:ScalarVectorBubble}, \protect\subref{fig:ScalarScalarBubble}, and  \protect\subref{fig:ScalarFermion} represent corrections to the scalar current. }\label{fig:Diagrams}
\end{figure}

\begin{figure}[h!]
	\centering
	\subfloat[]{
		\begin{turn}{90}
			\begin{tikzpicture}[scale=2.50]
				\begin{feynman} [inline=a]
					\diagram [horizontal=a to b] {			
					};
					\coordinate (c) at (-1,0);
					\coordinate (b);
					\fill (c) circle (2pt);	
					\draw [photon](b) -- (c);
					\draw [scalar] (b) arc [start angle=0, end angle=-360, radius=0.5cm];		
				\end{feynman}
			\end{tikzpicture}
		\end{turn}
		\label{fig:ScalarFourPoint}
	}
	\subfloat[]{
		\begin{turn}{90}
			\begin{tikzpicture}[scale=2.50]
				\begin{feynman} [inline=a]
					\diagram [horizontal=a to b] {			
					};
					\coordinate (c) at (-1,0);
					\coordinate (b);
					\fill (c) circle (2pt);	
					\draw [photon](b) -- (c);
					\draw [photon] (b) arc [start angle=0, end angle=-360, radius=0.5cm];		
				\end{feynman}
			\end{tikzpicture}
		\end{turn}
		\label{fig:VectorFourPoint}
	}
	\subfloat[]{
		\begin{tikzpicture}[scale=2.50]
			\begin{feynman} [inline=a]
				\diagram [horizontal=a to b] {			
				};
				\coordinate (b) at (0,0); 
				\coordinate (a) at (1,0);
				\coordinate (c) at (0.5,-0.5);
				\fill (c) circle (2pt);	
				\node[crossed dot, minimum size=0.3cm] at (0.5,0.5);
				\draw (b) arc [start angle=180, end angle=0, radius=0.5cm];
				\draw [anti fermion] (b) arc [start angle=-180, end angle=-90, radius=0.5cm];
				\draw [fermion] (c) arc [start angle=-90, end angle=0, radius=0.5cm];			
			\end{feynman}
		\end{tikzpicture}
		\label{fig:FermionMass}
	}
	\subfloat[]{
		\begin{tikzpicture}[scale=2.50]
			\begin{feynman} [inline=a]
				\diagram [horizontal=a to b] {			
				};
				\coordinate (b) at (0,0); 
				\coordinate (a) at (1,0);
				\coordinate (c) at (0.5,-0.5);
				\fill (c) circle (2pt);	
				\node[crossed dot, minimum size=0.3cm] at (0.5,0.5);
				\draw [scalar](b) arc [start angle=180, end angle=0, radius=0.5cm];
				\draw [scalar] (b) arc [start angle=-180, end angle=-90, radius=0.5cm];
				\draw [scalar] (c) arc [start angle=-90, end angle=0, radius=0.5cm];			
			\end{feynman}
		\end{tikzpicture}
		\label{fig:ScalarMass}
	}
	\caption{Additional diagrams that contribute to the vector self-energy at next-to-leading order. Diagrams \protect\subref{fig:FermionMass} and \protect\subref{fig:ScalarMass} correspond to mass insertions, and diagrams \protect\subref{fig:ScalarFourPoint} and \protect\subref{fig:VectorFourPoint} vanish in Feynman gauge.}\label{fig:Diagrams2}
\end{figure}
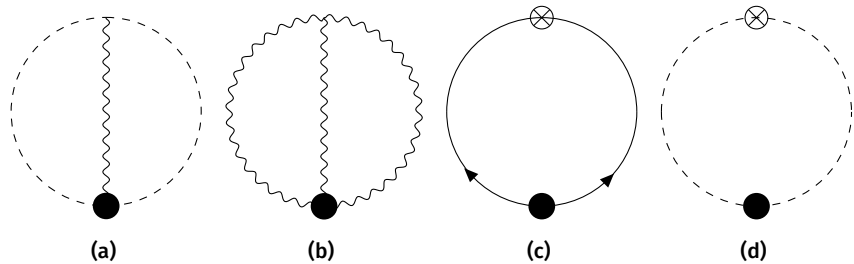

We also note that the calculation is simpler in Feynman gauge. In particular,  scalar and vector currents contain terms of the form
\begin{align}
 \bracket{A^a_\mu R^i R^j}, \quad  \bracket{A^a_\mu A^{b,\nu}A_\nu^c},
\end{align}
which at two loops give the diagrams shown in figures \ref{fig:ScalarFourPoint} and \ref{fig:VectorFourPoint}. However, in Feynman gauge these diagrams vanish. In addition, the ghost-current shown in figure \ref{fig:GhostVector} does not contribute at two loops in Feynman gauge.

\section{Generic models}\label{sec:Generic}
We denote scalar particles by $i,j,k,\ldots$; vector particles by $a,b,c,\ldots$; and fermions by $I,J,K,\ldots$. To parametrize a general model we use the Lagrangian~\cite{Machacek:1983tz,Machacek:1983fi,Machacek:1984zw,Martin:2018emo}
\begin{align}
	\mathcal{L}&=
	- \frac{1}{2} R_{i}^{ }(-\delta_{ij}^{ }\partial_\mu^{ } \partial^\mu + \mu_{ij}) R_{j}^{ }
	- \frac{1}{4}F_{\mu \nu}^a F^{\mu \nu,b}\delta_{ab}
	- \frac{1}{2 \xi_a}(\partial_\mu A^{a,\mu})^2 \nonumber
	\\ &
	- \partial^\mu \overline{\eta}^a \partial_\mu \eta^a
	+ i\psi^{\dagger,I}\overline{\sigma}^\mu \partial_\mu \psi_{I}
	- \frac{1}{2}(M^{IJ}\psi_{I}^{ } \psi_{J}^{ } + \text{h.c.})
	+ \mathcal{L}_\text{int} \nonumber
	\\
	\mathcal{L}_\text{int}&=
	- \frac{1}{4!}\lambda^{ijkm}R_{i}^{ } R_{j}^{ } R_{k}^{ } R_{m}^{ }
	- \frac{1}{2}(Y^{i I J }R_{i}^{ } \psi_{I}^{ } \psi_{J}^{ } + h.c)
	\\ &
	+ g^{a,I}_{J}A^a_\mu \psi^{\dagger,J}\overline{\sigma}^\mu \psi_{I}
	- g_{jk}^{a}A^a_\mu  R_{j}^{ } \partial^\mu R_{k}^{ } 
	- \frac{1}{2}g^{a}_{j n}g^b_{kn}A_\mu^a A^{\mu,b}R_{j}^{ } R_{k}^{ }
	- g^{abc}A^{\mu,a}A^{\nu,b}\partial_\mu^{ } A^c_\nu \nonumber
	 \\&
	- \frac{1}{4}g^{abe}g^{cde}A^{\mu a}A^{\nu b}A_{\mu}^c A_\nu^d
	+ g^{abc}A_\mu^a \eta^b \partial^\mu \overline{\eta}^c. \nonumber
\end{align}
In this notation $R_i$ are scalar fields in a real basis; $A^a_\mu$ are vector bosons; $\eta^a$ are ghosts; and $\psi_I$ are Weyl fermions~\cite{Dreiner:2008tw}. The sigma matrices are defined as
\begin{align}
	\sigma^{\mu}=\left(\mathbbm{1},\sigma^i\right),\quad
	\overline{\sigma}^{\mu}=\left(-\mathbbm{1},\sigma^i\right)
	\;,
\end{align}
and satisfy
\begin{align}
	\left\{\sigma_\mu,\ol{\sigma}_\nu \right\}=-2 g_{\mu \nu}, \quad g_{\mu\nu}=\text{diag}\left(-1,\vec{1}\right).
\end{align}
The couplings are normalized such that for the Standard-model we have
\begin{align*}
&\delta_{ab}\text{Tr}\left[g_V^a g_V^b\right]=-24 g_s^2-6 g_w^2,\quad  \delta_{ab}\text{Tr}\left[g_S^a g_S^b\right]=-3g_w^2-g_Y^2,
\\& \centering \delta_{ab} \text{Tr}\left[g_F^a g_F^b\right]=N_F\left(16 g_s^2+6 g_w^2+\frac{10}{3}g_Y^2\right).
\end{align*}
For a generic model these coupling tensors can be calculated by hand, but they are also straightforward to find from GroupMath~\cite{Fonseca:2020vke}.

To calculate hard thermal loops we use resummed distributions. For a general model these are~\cite{Blaizot:1999xk,Blaizot:1993zk}
\begin{align}
	&N_V^{\pm}\rightarrow N_V^{ab,\pm}=\delta^{ab} \left[\frac{1}{2}+n_B(p^0)\right]+ \delta N^{ab,\pm}_V(p^0,\vec{p}),
	\\&N_S^{\pm}\rightarrow N_S^{ij,\pm}=\delta^{ij} \left[\frac{1}{2}+n_B(p^0)\right]+ \delta N^{ij,\pm}_S(p^0,\vec{p}),
	\\&N_F^{\pm}\rightarrow N_{F,J}^{I,\pm}=\delta^{I}_J  \left[\frac{1}{2}-n_F(p^0)\right]- \delta N^{I,\pm}_{F,J}(p^0,\vec{p}),
\end{align}
where
\begin{align}
	n_B(p)=\left(e^{p/T}-1\right)^{-1}, \quad n_F(p)=\left(e^{p/T}+1\right)^{-1}.
\end{align}
We can condense the notation further:
\begin{align}
&  \delta N_V^{ab}\equiv -i g^{abc} \delta N_V^c, \quad \delta N_S^{ij}\equiv -i g^{c}_{ij} \delta N_S^c, \quad  \delta N_{F,J}^{I}\equiv  g^{c,I}_J \delta N_F^c,
\end{align}
where the distributions satisfy\footnote{We are for the moment omitting higher-point functions. These are calculated in section \ref{sec:HigherPoint}.}
\begin{align}
v\cdot \partial \delta N^{\pm,a}_X=\mp \vec{v}\cdot \vec{E}^{a} n'_X(p), \quad 
\vec{E}^a=-\dot{\vec{A}}^{a}-\vec{\nabla}A^{0,a}.
\end{align}

\newpage

\subsection{Conventions and structure of the calculation}\label{eq:TwoLoopRes}
All correlators that contribute at next-to-leading order are shown in figures \ref{fig:Diagrams} and \ref{fig:Diagrams2}, and the details are given in appendices \ref{app:VectorDiagrams}, \ref{app:FermionDiagrams}, \ref{app:ScalarDiagrams}, and \ref{app:CounterTerms}. We use
\begin{align}\label{eq:LorentzStruct}
	&\Pi_1^{\mu \nu}=\int\frac{d\Omega_v}{4\pi}\left(n^\mu n^\nu+k^0 \frac{v^\mu v^\nu}{v\cdot K}\right), \quad n^{\mu}=\left(1,\vec{0}\right)
	\\&  \Pi_2^{\mu \nu}=\int \frac{d\Omega_v}{4\pi}\left\{v^\mu v^\nu \left[\frac{(k^0)^2}{(v\cdot K)^2}-\frac{2 k^0}{v\cdot K}\right]+\left[v^\mu n^\nu+n^\mu v^\nu\right]\frac{ k^0}{v\cdot K}-n^\mu n^\nu\right\},
\end{align}
to signify the two Lorentz structures that appear. Note that these satisfy $K_\mu \Pi^{\mu \nu}=0$, so the self-energy is automatically transverse.

To derive the self-energy we need various currents:
\begin{align}
\partial_\mu F^{\nu\mu,a}= j^{a,\nu}_F+ j^{a,\nu}_S+ j^{a,\nu}_g+ j^{a,\nu}_V.
\end{align}
The fermion current is given by
\begin{align}
j^{a,\nu}_F=g^{a,J}_I \bracket{\psi^{\dagger,I}\overline{\sigma}^\nu \psi_J}.
\end{align}
The scalar current is
\begin{align}
j^{a,\nu}_S=\frac{1}{2!}g^a_{ij} \bracket{\partial^\nu R_i  R_j- R_i \partial^\nu R_j}.
\end{align}
The ghost current is
\begin{align}
j^{a,\nu}_g=g^{abc} \bracket{\eta^b \partial^\mu \overline{\eta}^c}.
\end{align}
Finally, the vector current is
\begin{align}
j^{a,\nu}_V=-g^{abc} \bracket{\partial_\mu A^{\nu,b}A^{\mu,c}+A^{\nu,b}\partial\cdot A^c+A_\mu^b \partial^\nu A^{\mu,c}-A^b\cdot \partial A^{\nu,c}}.
\end{align}

\subsection{One-loop hard thermal loops}\label{sec:HTLOneloop}
As mentioned, one-loop results are well known~\cite{Braaten:1989mz,Braaten:1990az,Blaizot:1993zk,Blaizot:1993be}. With our notation the results are
\begin{align}
	\Pi^{\mu \nu}_\text{LO}=\Pi^{\mu \nu}_V+\Pi^{\mu \nu}_F+\Pi^{\mu \nu}_S,
\end{align}
where
\begin{align}
	&\Pi^{\mu \nu}_V=-(D-2)\text{Tr}\left[g_V^a g_V^b\right]\int_p  n'_B(p) \Pi^{\mu \nu}_1=\frac{T^2}{3}\text{Tr}\left[g_V^a g_V^b\right]  \Pi^{\mu \nu}_1+\mathcal{O}(\epsilon),
	\\&\Pi^{\mu \nu}_F=2\text{Tr}\left[g_F^a g_F^b\right]\int_p n'_F(p) \Pi^{\mu \nu}_1=-\frac{T^2}{6}\text{Tr}\left[g_F^a g_F^b\right]  \Pi^{\mu \nu}_1+\mathcal{O}(\epsilon),
	\\&\Pi^{\mu \nu}_S=-\text{Tr}\left[g_S^a g_S^b\right]\int_p n'_B(p) \Pi^{\mu \nu}_1=\frac{T^2}{6}\text{Tr}\left[g_S^a g_S^b\right]  \Pi^{\mu \nu}_1+\mathcal{O}(\epsilon).
\end{align}

\subsection{Two-loop hard thermal loops}\label{sec:HTLTwoloop}
 At two loops various diagrams introduce factors of $D=4-2\epsilon$; below we have only kept the $\mathcal{O}(\epsilon^0)$ contribution, but the full results are given in the appendices.
We separate the result as
\begin{align}\label{eq:NLORes}
	\Pi^{\mu \nu,ab}_\text{NLO}=-\left[\Pi^{\mu \nu,ab}_\text{V}+\Pi^{\mu \nu,ab}_\text{SV}+\Pi^{\mu \nu,ab}_\text{FV}+\Pi^{\mu \nu,ab}_\text{SF}\right],
\end{align}
signifying pure vector, scalar-vector, fermion-vector, and scalar-fermion-vector type interactions respectively. All repeated indices are summed.

Let us start with the pure-vector contribution:
\begin{align}\label{eq:TwoLoop}
	&\Pi^{\mu \nu,ab}_\text{V}=11  T^2 \frac{ \log (\frac{\mu e^\gamma}{4 \pi T} ) -\frac{1}{22}}{36 \pi ^2}g_V^{adc}g_V^{cef}g_V^{dfn}g_V^{enb}\Pi_1^{\mu \nu}-\frac{T^2}{12 \pi ^2} g_V^{adc}g_V^{cef}g_V^{dfn}g_V^{enb} \Pi_2^{\mu \nu}.
\end{align}

The scalar-vector contribution is
\begin{align}
	&\Pi^{\mu \nu, ab}_\text{SV}=\frac{T^2}{192 \pi ^2} \text{Tr}\left[g_S^a g_S^b\right]_{jl}\lambda^{jlnn} \Pi_2^{\mu \nu}+\frac{1}{8 \pi ^2}\text{Tr}\left[g_S^a g_S^b\right]_{ij}\mu^{ij} \Pi_2^{\mu \nu} \nonumber
	\\& -T^2 \frac{ \log \frac{\mu e^\gamma}{4 \pi T}  +1}{288 \pi ^2}\text{Tr}\left[g_S^a g_S^c\right]\text{Tr}\left[g_S^c g_S^b\right]\Pi_1^{\mu \nu}
	- \frac{T^2}{32 \pi ^2}\text{Tr}\left[g_S^a g_S^b g_S^c g_S^c\right] \Pi_2^{\mu \nu}\nonumber
	\\&-T^2 \frac{\log\frac{\mu e^\gamma}{4\pi T}}{24\pi^2} g _V^{aec}g_V^{bdc}\text{Tr}\left[g_S^d g_S^e\right] \Pi_1^{\mu \nu}+ \frac{T^2}{48\pi^2}g _V^{aec}g_V^{bdc}\text{Tr}\left[g_S^d g_S^e\right]\Pi_2^{\mu \nu} \nonumber
	\\&+T^2\frac{ \log \frac{\mu e^\gamma}{4\pi T}-3}{72 \pi ^2}\left\{\text{Tr}\left[g_S^a g_S^c\right]\text{Tr}\left[g_V^c g_V^b\right] +\text{Tr}\left[g_V^a g_V^c\right]\text{Tr}\left[g_S^c g_S^b\right]  \right\}\Pi_1^{\mu \nu}.
\end{align}

The fermion-vector contribution is
\begin{align}
	&	\Pi^{\mu \nu, ab}_\text{FV}= T^2 \frac{  \log \frac{\mu e^\gamma}{4 \pi T} }{24 \pi ^2} g_V^{ace}g_V^{bcd}\text{Tr}\left[g_F^d g_F^e\right]\Pi_1^{\mu \nu}\nonumber
	-T^2  \frac{ \log \frac{\mu e^\gamma}{4\pi T} -\frac{1}{2}+\log (4)}{72 \pi ^2} \text{Tr}g_F^a g_F^c\text{Tr}g_F^c g_F^b\Pi_1^{\mu \nu}\nonumber
	\\& -T^2\frac{ \log \frac{\mu e^\gamma}{4\pi T} +\frac{3}{2}-8 \log (2)}{288 \pi ^2} \left\{\text{Tr}\left[g_F^a g_F^c\right]\text{Tr}\left[g_V^c g_V^b\right] +\text{Tr}\left[g_V^a g_V^c\right]\text{Tr}\left[g_F^c g_F^b\right]  \right\}\Pi_1^{\mu \nu}\nonumber
	\\&+\frac{T^2}{16 \pi ^2}\text{Tr} g_F^c g_F^c g_F^a  g_F^b \Pi_2^{\mu \nu}
	-\frac{T^2}{48 \pi ^2}g_V^{ace}g_V^{bcd}\text{Tr}\left[g_F^d g_F^e\right]\Pi_2^{\mu \nu} +\frac{1}{8 \pi ^2}\text{Tr}\left[g_F^a M_F M_F^{\dagger} g_F^b\right]\Pi_2^{\mu \nu}.
\end{align}

And finally, the mixed fermion-scalar contribution is
\begin{align}
	&\Pi^{\mu \nu,ab}_\text{SF}=T^2\frac{5 \log \frac{\mu e^\gamma }{4 \pi  T} -1+8 \log (2)}{576 \pi ^2}\left\{\text{Tr}\left[g_S^a g_S^c\right]\text{Tr}\left[g_F^c g_F^b\right] +\text{Tr}\left[g_F^a g_F^c\right]\text{Tr}\left[g_S^c g_S^b\right]  \right\} \Pi_1^{\mu \nu}\nonumber
	\\&+\frac{T^2}{32\pi^2}\left[g_F^a g_F^b\right]^{I}_J (Y Y^c)^{J}_I\Pi_2^{\mu \nu}
	+\frac{T^2}{192\pi^2}\left[g_S^a g_S^b\right]_{ij}(Y Y^c+Y^c Y)^{ij}\Pi_2^{\mu \nu}.
\end{align}

In the traces over generators the contractions are made with the conventions
\begin{align}
\text{Tr}\left[g_V^a g_V^b\right]=g^{acd} g^{bdc}, \quad \text{Tr}\left[g_S^a g_S^b\right]=g^a_{ij} g^b_{ji}, \quad \text{Tr}\left[g_F^a g_F^b\right]=g^{a,I}_J g^{b,J}_I.
\end{align}

Note that our two-loop results in equation \eqref{eq:TwoLoop} ensures that
\begin{align}
		\Pi^{\mu \nu}=	\Pi^{\mu \nu}_\text{LO}+\Pi^{\mu \nu}_\text{NLO},
\end{align}
is renormalization-scale invariant. As such one should choose $\mu\sim T$ to ensure that no large logarithms are present.

\subsection{Power corrections from one-loop diagrams}\label{sec:PowerCorrection}
Power corrections modify the kinetic terms and are, for example, responsible for anomalous dimensions.
 We forgo using transport equations since the diagrams are straightforward to evaluate~\cite{Gorda:2022fci,Carignano:2019ofj,Carignano:2021zhu}. 
 
We use a convention where the Debye mass is given by 
\begin{align}
	(m_D^2)^{ab}=-\lim_{k^0\rightarrow 0}\Pi^{\mu \nu,ab}_\text{NLO},
\end{align}
with $\Pi^{\mu \nu;ab}_\text{NLO}$ defined by equation \eqref{eq:NLORes}. This means that we have rescaled our vector fields to make the $A^{0}$ kinetic term canonical when $k^0=0$. To wit, we have moved all renormalization-scale dependence (and some finite pieces) away from the power corrections\footnote{Our convention makes the Lorentz structure easy at two loops, in addition, the result is manifestly renormalization-scale invariant.}. The original results\te before field-redefinitions\te are given in appendix \ref{app:PowerFull}.

That said, the scalar loop gives
\begin{align}
\Pi^{\mu \nu,ab}_{S}(K)=-\text{Tr}\left[g_S^a g_S^b\right] \int_P \left(2 P+K\right)^\mu\left(2 P+K\right)^\nu \Delta_S(P) \Delta^R(P+K).
\end{align}
We are only interested in the sub-leading correction scaling as $K^2\sim k^2\sim (g T)^2$. After expanding the integral, and adding counter-terms, we find
\begin{align}
&g_{\mu \nu }\Pi^{\mu \nu,ab}_{S}(K)=\text{Tr}\left[g_S^a g_S^b\right]\frac{K^2}{16 \pi^2}\left\{-\frac{2}{3}+k^0 L(K)\right\},
\\& \Pi^{00,ab}_S(K)=-\text{Tr}\left[g_S^a g_S^b\right]\frac{k^2}{16 \pi^2}\left\{\frac{1}{3} \frac{(k^0)^2}{k^2} (k^0 L(K)-1)\right\}
\end{align}

The fermion loop gives
\begin{align}
	&\Pi^{\mu \nu,ab}_{F}(K)=\text{Tr}\left[g_F^a g_F^b\right] \int_P F^{\mu \nu} \Delta_F(P) \Delta^R(P+K),
	\\&F^{\mu \nu}=2\left[-g^{\mu \nu}\left(K\cdot P+P^2\right)+2 p^\mu p^\nu +k^\mu p^\nu+k^\nu p^\mu\right].
\end{align}
After expanding the integral, and adding counter-terms, we find
\begin{align}
	&g_{\mu \nu }\Pi^{\mu \nu,ab}_{F}(K)=\text{Tr}\left[g_F^a g_F^b\right]\frac{K^2}{16 \pi^2}\left\{\frac{4}{3}+4 k^0 L(K)\right\},
	\\& \Pi^{00,ab}_F(K)=-\text{Tr}\left[g_F^a g_F^b\right]\frac{k^2}{16 \pi^2}\left\{\frac{2}{3} k^0 \left(3-\frac{(k^0)^2}{k^2}\right)L(K)+\frac{2}{3}\frac{(k^0)^2}{k^2}\right\}.\nonumber
\end{align}

For non-abelian diagrams we group ghosts and vectors together. After adding counter-terms we find
\begin{align}
	&g_{\mu \nu }\Pi^{\mu \nu,ab}_{V}(K)=\text{Tr}\left[g_V^a g_V^b\right]\frac{K^2}{16 \pi^2}\left\{\frac{4}{3}+10 k^0 L(K)\right\},
	\\& \Pi^{00,ab}_V(K)=-\text{Tr}\left[g_V^a g_V^b\right]\frac{k^2}{16 \pi^2}\left\{\frac{2}{3} k^0 \left(6-\frac{(k^0)^2}{k^2}\right)L(K)+\frac{2 (k^0)^2}{3 k^2}\right\}.\nonumber
\end{align}

\subsection{Transverse and longitudinal self-energies}\label{sec:TransLong}
It is useful to write the vector self-energy in terms of transverse and longitudinal components~\cite{Pisarski:1989cs}:
\begin{align}
&\Pi^{\mu \nu}=\Pi_T P_T^{\mu \nu}+\Pi_L P_L^{\mu \nu}, \quad &P_T^{i j}=\delta^{ij}-\frac{p^i p^j}{p^2}, \quad P_L^{\mu \nu}=g^{\mu \nu}-\frac{K^\mu K^\nu}{K^2}-P_T^{\mu \nu}.
\end{align}
We then find
\begin{align}
\Pi_T=\frac{1}{d-1}\left[g_{\mu \nu}\Pi^{\mu \nu}+\frac{K^2}{k^2}\Pi^{00}\right], \quad \Pi_L=-\frac{K^2}{k^2}\Pi^{00}.
\end{align}
Since our results are built from the Lorentz structures $\Pi_1^{\mu\nu}$ and $\Pi_2^{\mu\nu}$ defined in equation \ref{eq:LorentzStruct}, we only need to find the traces of these. To wit
\begin{align}
&\Pi_1^{00}=1- k^0 L[K], \quad g_{\mu \nu }\Pi_1^{\mu \nu}=-1,
\\&\Pi_2^{00}=-1-\frac{(k^0)^2}{K^2}, \quad g_{\mu \nu }\Pi_2^{\mu \nu}=1+2 k^0 L[K],
\\& L[K]\equiv \frac{1}{2k}\log \frac{k^0+k+i \eta}{k^0-k+\eta}, \quad \eta=0^{+},
\end{align}
where we have used known results for the angular integrals~\cite{Laine:2016hma,Gorda:2022fci}.

\subsection{Examples}
Consider now the gluon self-energy with $N_q$ quarks:
\begin{align}
	\Pi_\text{NLO}^{\mu \nu}&=\frac{g_s^4 (N_q+6) T^2 \left[(4 N_q-66) \log \frac{\mu e^\gamma}{4\pi T}-2 N_q+8 N_q \log (2)+3\right]}{288 \pi ^2} \Pi_1^{\mu \nu}
	\\&-\frac{g_s^4 (N_q-18) T^2}{48 \pi ^2} \Pi_2^{\mu \nu}.
\end{align}

Next, the Standard-Model. The gluon self-energy is
\begin{align}
	\Pi_\text{NLO}^{\mu \nu}&=-\frac{g_s^4 T^2 \left[14 \log \frac{\mu e^\gamma}{4\pi T}+3-16 \log (2)\right]}{8 \pi ^2} \Pi_1^{\mu \nu}
	\\&-\frac{g_s^2 T^2 \left[-48 g_s^2+27 g_w^2+11 g_Y^2+12 y_t^2\right]}{192 \pi ^2}\Pi_2^{\mu \nu}.
\end{align}
Here $g_s$ is the strong coupling constant, $g_w$ the weak one, $g_Y$ is the hypercharge coupling, and $y_t$ is the top-Yukawa coupling.

Finally, take an $\mathrm{SO}(10)$ gauge theory with $N_F$ fermions in the spinor ($16$) representation, and a $45\oplus 16$ Higgs. The gauge self-energy is
\begin{align}
	\Pi_\text{NLO}^{\mu \nu}&=\frac{g_x^4 (N_F+14) T^2 \left[(2 N_F-41) \log \frac{\mu e^\gamma}{4\pi T}+N_F (\log (16)-1)+5\right]}{36 \pi ^2} \Pi_1^{\mu \nu}
	\\&-\frac{g_x^4 (71 N_F-1415) T^2}{192 \pi ^2} \Pi_2^{\mu \nu}.
\end{align}

\section{Higher-point hard thermal loops}\label{sec:HigherPoint}
So far we have focused on the self-energy, but higher-point correlators can be extracted from the results in section \ref{sec:HTLTwoloop}. In particular, it is well-known that at one loop all higher-point functions can be derived by using~\cite{Blaizot:1993be,Blaizot:1993zk,Blaizot:1999xk}
\begin{align}
\left[v\cdot D, \delta N^{\pm}_X\right]^a=\mp \vec{v}\cdot \vec{E}^{a} n'_X(p),
\end{align}
where the covariant derivative is $\left[ D_\mu N\right]^a=\partial_\mu N^a+g^{abc}A_\mu^b N^c$. We can then expand the currents as
\begin{align}
j^{a}_\mu=\Pi^{ab}_{\mu\nu}A^{\nu,b}+\frac{1}{2}\Gamma_{\mu\nu\rho}^{abc}A^{\nu,b} A^{\rho,c}+\ldots
\end{align}
To find these higher-point functions we can use the results in section \ref{eq:TwoLoopRes} together with the replacements:
\begin{align}
&C^{ab}\Pi_1^{\mu \nu}\rightarrow C^{ab} \int \frac{d\Omega_v}{4\pi} \left[\frac{v^\mu \vec{v}\cdot \vec{E}}{v\cdot D}\right]^b,
\\& D^{ab}\Pi_2^{\mu \nu}\rightarrow D^{ab}  \int \frac{d\Omega_v}{4\pi}  \left\{v^\mu\left(-\frac{D_0}{(v\cdot D)^2}-\frac{1}{v\cdot D}\right) \vec{v}\cdot \vec{E}-\frac{v^\mu-n^\mu}{v\cdot D}\vec{v}\cdot \vec{E}\right\}^b,
\end{align}
where now $E^{a,i}=\partial^i A^{0,a}-\partial^0 A^{i,a}+g^{abc}A^{i,b}A^{0,c}$. 

Consider the first Lorentz-structure, which coincides with the one-loop one. The corresponding three-point vertex is well-known~\cite{Andersen:2002ey,Braaten:1989mz,Blaizot:1993be}:
\begin{align}\label{eq:ThreePoint}
&C^{ab}\Pi_1^{\mu \nu}\rightarrow -iC^{ae}g^{ebc} \Gamma_1^{\mu\nu\rho}(P,Q,R), \quad \Gamma_1^{\mu\nu\rho}(P,Q,R) = \int \frac{d\Omega_v}{4\pi} \frac{v^\mu v^\nu v^\rho}{v\cdot P}\left[\frac{q^0}{v\cdot Q}-\frac{r^0}{v\cdot R}\right].
\end{align}

Likewise, it is possible to find the three-point vertex corresponding to $\Pi_2^{\mu \nu}$ by expanding the covariant derivatives. Yet it is easier to exploit that this new Lorentz structure arises because the ballistic approximation ceases to hold: 
\begin{align}
	v_p^\mu \rightarrow v_p^\mu -\frac{m^2}{2 p^2}(v^\mu_p-n^\mu)+\ldots
\end{align}
As such we can use equation \ref{eq:ThreePoint}\te together with the correction above\footnote{We have to remember that the original term depends on $\int dp p^2 n'(E)$, which when $E\approx p+\frac{m^2}{2 p}$ becomes $\int dp \left[p^2 n'(p)-\frac{m^2}{2}n'(p)\right]$. }\te and collect all terms proportional to $m^2$:
\begin{align}
D^{ab}\Pi_2^{\mu \nu}&\rightarrow -i D^{ae}g^{ebc} \Gamma^{\mu \nu \rho}_2(P,Q,R),
\\&\Gamma^{\mu \nu \rho}_2(P,Q,R)= \int \frac{d\Omega_v}{4\pi} \frac{-2 v^\mu v^\nu v^\rho+\left(n^\mu v^{\nu}v^\rho +\text{perm}\right)}{v\cdot P}\left[\frac{q^0}{v\cdot Q}-\frac{r^0}{v\cdot R}\right]
\\& +\int \frac{d\Omega_v}{4\pi} \frac{ v^\mu v^\nu v^\rho}{v\cdot P}\left[\frac{p^0}{v\cdot P}\left(\frac{q^0}{v\cdot Q}-\frac{r^0}{v\cdot R}\right)+\left(\frac{(q^0)^2}{(v\cdot Q)^2}-\frac{(r^0)^2}{(v\cdot R)^2}\right)\right].
\end{align}

Note that the Ward identity is automatically satisfied since
\begin{align}
& P_\mu \Gamma_1^{\mu \nu \rho}(P,Q,R)=\Pi_1^{\nu \rho}(Q)-\Pi_1^{\nu \rho}(R),
\\&P_\mu \Gamma_2^{\mu \nu \rho}(P,Q,R)=\Pi_2^{\nu \rho}(Q)-\Pi_2^{\nu \rho}(R).
\end{align}

The same procedure can be applied to four-point interactions, which at one-loop are given in~\cite{Andersen:2002ey,Braaten:1989mz}.

\section{Conclusions}

In this paper we have provided hard thermal loops, for vector boson self-energies, at two-loops for any renormalizable model. This was made possible by using transport equations to simplify the calculations\te thus extending known one-loop methods~\cite{Blaizot:1993zk,Blaizot:1993be,Litim:2001db,Blaizot:2001nr}. In particular, this approach provides compact expression for each particle type; the result is independent of the matching scale; and known results for Debye masses are reproduced in the appropriate limit. We also demonstrated how higher-point functions can be extracted from the results.

The results of this paper can be used to study particle production in the early universe; transport coefficients; and wall speeds in first-order phase transitions~\cite{Moore:1995si}. The effect of including two-loop contributions is likely significant for the strong interaction, since the coupling constant is rather large $N \alpha_S\sim 0.3$ when $T\sim 100~\text{GeV}$.

The next step is to provide two-loop hard thermal loops for fermion propagators. Performing these calculations for quarks, by using Feynman diagrams, is likely arduous beyond leading order. However, we expect that similar methods as used in this paper will prove useful in this endeavour. 

\section*{Acknowledgements}
I am grateful to the high-energy physics group at the University of Granada for their hospitality as this work was being finished. I also want to thank Geraldine Servant for help with the manuscript.
This work has been supported by the Swedish Research Council, project number VR:$2021$-$00363$ and by the Deutsche Forschungsgemeinschaft under Germany’s Excellence Strategy - EXC $2121$ Quantum Universe - $390833306$.

\appendix

\section{Derivatives of resummed distributions } \label{app:DerivDistri}
The distributions satisfy 
\begin{align}
 \delta N^{\pm}_X=\mp \frac{ e p_\alpha F^{\alpha \beta}}{p\cdot \partial}\partial^\beta_p n_X(p^0).
\end{align}
So taking the derivative $\frac{\partial}{\partial p^0}$ and going to momentum space we find
\begin{align}
	\int p dp	\frac{\partial}{\partial p^0 }\delta N^{\pm}_X\rightarrow \mp \int dp \left[\frac{k^0 }{(v\cdot K)^2}-\frac{1}{v \cdot K}\right]\vec{v}\cdot \vec{E}(K) n'_X(p).
\end{align}

\subsection{Momentum integrals}\label{app:RadialIntegral}
We use dimensional regularization where $d=3-2\epsilon$. When evaluating the self-energy we encounter the integrals
\begin{align}
&T^{2\epsilon}\int dp p^{d-1} n'_B(p)=-\frac{1}{3} \pi ^2 T^2+\frac{1}{3} \pi ^2 T^2 (-24 \log (A)+3+\log (4)+2 \log (\pi ))\epsilon,
\\& T^{2\epsilon}\int dp p^{d-1} n'_F(p)=-\frac{1}{6} \pi ^2 T^2+ \frac{1}{6} \pi ^2 T^2 (-24 \log (A)+3+\log (16)+2 \log (\pi ))\epsilon,
\\&T^{2\epsilon} \int dp p^{d-3} n_B(p)=-\frac{T}{2 \epsilon}+\mathcal{O}\left(\epsilon\right), \quad T^{2\epsilon} \int dp p^{d-3} n_F(p)=T \log (2)+\mathcal{O}\left(\epsilon\right),
\\& T^{2\epsilon} \int dp p^{d-3} n'_B(p)=\frac{1}{2} +\mathcal{O}\left(\epsilon\right), \quad T^{2\epsilon} \int dp p^{d-3} n'_F(p)=-\frac{1}{2}+\mathcal{O}\left(\epsilon\right)
\end{align}
Here $\text{A}\approx 1.28243$ is the Glaisher constant.

\section{Non-abelian gauge theories}\label{app:VectorDiagrams}
Note that all (collinear) divergences resulting from angular integrations cancel. We will however obtain divergences\te real ones\te from radial integrations: $\int_p \frac{n_B(p)}{p^2}\sim -\frac{T}{2 \epsilon}$. The $\epsilon$ poles from these terms cancel once zero-temperature counterterms are used.

Throughout this and the following sections we  keep factors of $D=4-2\epsilon$ explicit. There are four contributions. Corrections to the vector current are shown in figures \ref{fig:VectorVector}, \ref{fig:VectorGhost}, and \ref{fig:VectorVectorBubble}; sunset corrections to the ghost current are shown in figure \ref{fig:GhostVector}. We also note that diagram \ref{fig:VectorFourPoint} vanishes.

\subsection{Vector current}
We start with the vector current. The sunset diagram gives
\begin{align}\label{eq:QCDGluon}
\Pi_{\ref{fig:VectorVector}}^{\mu}=-\frac{1}{4}g^{abc}g^{hgn}g^{def}\int_{PQ} &F^{\mu}\left\{\delta^{en}\delta^{cd}\Delta_V^{bh}(P)\Delta_V^{gf}(Q)\left[\Delta^R(P)\Delta^R(P+Q)+\Delta^A(P)\Delta^A(P+Q)\right]\right.
\nonumber\\&\left. \delta^{cd}\delta^{gf}\Delta_V^{bh}(P)\Delta_V^{en}(P+Q)\left[\Delta^R(P)\Delta^A(Q)+\Delta^A(P)\Delta^R(Q)\right] \right. \nonumber
\\& \left. \delta^{cd}\delta^{bh}\Delta_V^{gf}(Q)\Delta_V^{en}(P+Q)\Delta^A(P)\Delta^R(P) \right\}
\end{align}
where
\begin{align}
F^\mu(P,Q)=&(P+Q)^2\left[(5-4D)p^\mu+	2(2D-3) q^\mu\right]+P^2\left[(5-6D)p^\mu\right] \nonumber
\\&+Q^2 \left[(11-8D)p^\mu+2(3-2D)q^\mu\right].
\end{align}
We can rewrite the terms so that they all multiply $\Lambda^{ab}=g_V^{anc}g_V^{cef}g_V^{nfn}g_V^{enb}$. Explicitly,
\begin{align}
&\Pi_{\ref{fig:VectorVector}}^{\mu}=\Lambda^{ab} \frac{(4D-5)}{4} \int_{PQ} n_B(q)\frac{1}{p q}\left\{\partial^p_0\left[N(p)_V-\overline{N}_V(p)\right]v^\mu_p-\frac{v^\mu_p-n^\mu}{p}\left(N_V(p)-\overline{N}_V(p)\right)\right\}^{b}  \nonumber
\\&+ \Lambda^{ab} \frac{(2D-3)}{4} \int_{PQ}v^\mu_p \frac{1}{q^2}\left(n_B(q)-q n_B'(q)\right)\left(N_V(p)-\overline{N}_V(p)\right)^{b} 
\end{align}

We now turn to the bubble diagram. We find
\begin{align*}
&\Pi_{\ref{fig:VectorVectorBubble}}^\mu=\frac{1}{4}g_V^{abl}g_V^{ecg}g_V^{fdg}\delta^{dl}\int_{PQ}F^\mu \Delta_V^{bc}(P)\left(\Delta^R(P)+\Delta^A(p)\right)\Delta_V^{fg}(Q), \quad F^\mu=- 4(D-1)^2p^\mu.
\end{align*}
The result is
\begin{align}
&\Pi^\mu_{\ref{fig:VectorVectorBubble}}=	-\frac{(D-1)^2}{2}\Lambda^{ab}\int_{PQ} n_B(q)\frac{1}{p q}\left\{\partial^p_0\left[N_V(p)-\overline{N}_V(p)\right]v^\mu_p-\frac{v^\mu_p-n^\mu}{p}\left(N_V(p)-\overline{N}_V(p)\right)\right\}^{b}.\nonumber
\end{align}

\subsection{Ghost diagrams}
We now consider diagrams with internal ghosts. There are two contributions: The ghost-current with an internal vector and the vector current with a ghost loop\te shown in figures \ref{fig:GhostVector} and \ref{fig:VectorGhost} respectively. The latter diagram vanish, so we only need the former one:
\begin{align}\label{eq:QCDGhost1}
	\Pi_{\ref{fig:VectorGhost}}^{\mu}=\frac{1}{2}g_V^{abc}g_V^{hgn}g_V^{def}\int_{PQ} &F^{\mu}\left\{\delta^{en}\delta^{cd}\Delta_V^{bh}(P)\Delta_V^{,gf}(Q)\left[\Delta^R(P)\Delta^R(P+Q)+\Delta^A(P)\Delta^A(P+Q)\right]\right.
	\nonumber\\&\left. \delta^{cd}\delta^{gf}\Delta_V^{bh}(P)\Delta_V^{en}(P+Q)\left[\Delta^R(P)\Delta^A(Q)+\Delta^A(P)\Delta^R(Q)\right] \right. \nonumber
	\\& \left. \delta^{cd}\delta^{bh}\Delta_V^{gf}(Q)\Delta_V^{en}(P+Q)\Delta^A(P)\Delta^R(P) \right\},
	\\& F^\mu=(P+Q)^2(q^\mu-p^\mu/2)+P^2 p^\mu/2-Q^2(q^\mu+\frac{3}{2}p^\mu). \nonumber
\end{align}
We find
\begin{align}
	\Pi_{\ref{fig:VectorGhost}}^{\mu}=&-\Lambda^{ab} \frac{1}{4}\int_{PQ}\frac{ n_B(q)}{p q}\left\{\partial^p_0\left[N_V(p)-\overline{N}_V(p)\right]v^\mu_p-\frac{v^\mu_p-n^\mu}{p}\left(N_V(p)-\overline{N}_V(p)\right)\right\}^{b}\nonumber
\\&-\frac{1}{4} \Lambda^{ab}  \int_{PQ}v^\mu_p \frac{n_B(q)-q n_B'(q)}{q^2}\left(N_V(p)-\overline{N}_V(p)\right)^{b}.
\end{align}

\subsection{Total contribution from non-abelian diagrams}
We find
\begin{align}
\Pi^{\mu\nu}_{\ref{fig:VectorVector}}+\Pi^{\mu\nu}_{\ref{fig:VectorVectorBubble}}+\Pi^{\mu\nu}_{\ref{fig:VectorGhost}}&=T^2 \frac{(D-2)^2}{48 \pi^2}\Lambda^{ab} \Pi_2^{\mu \nu}-2(D-2)\Lambda^{ab} I_\text{VV}\Pi^{\mu \nu}_1,
\end{align}
where
\begin{align}
I_\text{VV}=\int_{PQ} n'_B(q)\frac{1}{p^2}\left(n'_B(p)-\frac{n_B(p)}{p}\right)=T^2\left\{\frac{1}{48 \pi^2 \epsilon}+\frac{(24 \log (A)+4 \log \frac{\mu}{4\pi T}+2 \gamma -1)}{48 \pi ^2}\right\},\nonumber
\end{align}
and $\Lambda^{ab}=g_V^{anc}g_V^{cef}g_V^{nfn}g_V^{enb}$.

\section{Fermion diagrams}\label{app:FermionDiagrams}
\subsection{Fermion current}
We now turn diagrams with fermions. We will omit collinear divergences as they cancel once we sum fermion and vector currents. The fermion current gives
\begin{align}\label{eq:FermionCurrentGen}
	\Pi_{\ref{fig:FermionVector}}^\mu=g^{a N}_I g^{c J}_K g^{d L}_M\int_{PQ} F^{\mu}(P,Q)&\left\{ \delta^{M}_N \delta^K_L \Delta_{F,J}^{I}(P)\Delta_V^{cd}(Q) \left[ \Delta^R(P)\Delta^R(P+Q)+\Delta^A(P)\Delta^A(P+Q)\right]\right.\nonumber
	\\&\left. \delta^{ac} \delta^{M}_N\Delta_{F,J}^{I}(P)\Delta^{K}_{F,L}(P+Q)\left[\Delta^R(P)\Delta^A(Q)+\Delta^A(P)\Delta^R(Q)\right] \right. \nonumber
	\\& \left.  \delta^{M}_N \delta^{I}_J \Delta^{K}_{F,L}(P+Q)  \Delta^{ac}_V(Q)\left[\Delta^R(P)\Delta^A(P)\right]\right\},\nonumber
	\\& F^\mu=-2(D-2)\left[(P+Q)^2 p^\mu-P^2 q^\mu -Q^2 p^\mu\right]
\end{align}
We find
\begin{align}
	&\Pi_{\ref{fig:FermionVector}}^\mu=\frac{D-2}{2} \text{Tr} g_F^c g_F^c g_F^a  g_F^b  \int_{PQ}\frac{N_B(q)-N_F(q)}{pq}\left\{\partial^p_0\left[N_F^{+}(p)-N_F^{-}(p)\right]v^\mu_p-\frac{v^\mu_p-n^{\mu}}{p}(N_F^{+}(p)-N_F^{-}(p))\right\} ^b. \nonumber
\end{align}
Here we should use the leading-order relation: $N_B(q)-N_F(q)=n_B(q)+n_F(q)$.
After inserting the resummed propagators and performing the integrals we find
\begin{align}
	\Pi_{\ref{fig:FermionVector}}^{\mu \nu}=-\frac{(D-2)T^2}{32 \pi ^2}\text{Tr} g_F^c g_F^c g_F^a  g_F^b \Pi_2^{\mu \nu}.
\end{align}

\subsection{Vector current}
Consider now fermion corrections to the vector current:
\begin{align}
	\Pi_{\ref{fig:VectorFermion}}^\mu=-\frac{1}{2}g_V^{ace}g^{d,I}_J g^{f,K}_L &\int_{PQ} F^{\mu} \delta^J_K \delta^{ef} \Delta_V^{cd}(P)\Delta_{F,I}^{L}(Q)\left[\Delta^R(P)\Delta^R(P+Q)+\Delta^A(P)\Delta^A(P+Q)\right]\nonumber
	\\&+\delta^I_L \delta^{ef} \Delta_V^{cd}(P)\Delta_{F,K}^{J}(P+Q)\left[\Delta^R(P)\Delta^A(Q)+\Delta^A(P)\Delta^R(Q)\right]\nonumber
	\\&+\delta^{ef}\delta^{cd}\Delta_{F,K}^J(P+Q)\Delta^{L}_{F,I}(Q)\left[\Delta^R(P)\Delta^A(P)\right],
\end{align}
where
\begin{align}
F^\mu&=-2 i\left\{(P+Q)^2 \left[ (D-2)p^\mu+2 q^\mu\right]-  (D-2)P^2 p^\mu+Q^2 \left[(D-4)p^\mu-2 q^\mu\right]\right\}.
\end{align}
We are left with
\begin{align}
&\Pi_{\ref{fig:VectorFermion}}^{\mu}=-\frac{(D-2)}{2} g_V^{ace}g_V^{bcd}\text{Tr}\left[g_F^d g_F^e\right]\int_{PQ}\frac{n_F(q)}{q p}\left\{\partial^p_0\left[N_V(p)-\overline{N}_V(p)\right]v^\mu_p-\frac{(v^\mu_p-n^\mu)}{p}\left(N_V(p)-\overline{N}_V(p)\right)\right\}^{b} \nonumber
\\& +i g_V^{ace}\text{Tr}\left[(g_F^c g_F^e-g_F^e g_F^c)g_F^b\right]\int_{PQ}(N_F(q)-\overline{N}_F(q))^b \frac{v^\mu_q}{p^2}\left\{n'_B(p)-\frac{n_B(p)}{p}\right\}
\end{align}
This result can be further simplified because $-ig_V^{ace}\text{Tr}\left[(g_F^c g_F^e-g_F^e g_F^c)g^b_F\right]=g_V^{ace}g_V^{bcd}\text{Tr}\left[g_F^d g_F^e\right]$, so the entire diagram is proportional to the structure $g_V^{ace}g_V^{bcd}\text{Tr}\left[g_F^d g_F^e\right]$. In any case, after inserting the resummed propagators and performing the integrals we find
\begin{align}
\Pi_{\ref{fig:VectorFermion}}^{\mu \nu}&=(D-2) \frac{T^2}{96 \pi^2}g_V^{ace}g_V^{bcd}\text{Tr}\left[g_F^d g_F^e\right] \Pi^{\mu \nu}_2+T^2g_V^{ace}g_V^{bcd}\text{Tr}\left[g_F^d g_F^e\right] I_\text{FV}\Pi^{\mu \nu}_1,
\end{align}
where
\begin{align*}
 I_\text{FV}=\int_{PQ} n'_F(q)\frac{1}{p^2}\left(n'_B(p)-\frac{n_B(p)}{p}\right)=T^2\left\{\frac{1}{48 \pi^2 \epsilon}-\frac{- 24 \log A-4 \log \frac{\mu}{4\pi T}-2 \gamma +1+\log (4)}{48 \pi ^2}\right\},
\end{align*}
contains divergences that cancel against counter-term insertions.

\subsection{Yukawa diagrams}
There are two diagrams with Yukawa couplings, one from the fermion current and one from the scalar current. The sum of the two gives
\begin{align}
\Pi_{\ref{fig:ScalarFermion}}^{\mu \nu}+\Pi_{\ref{fig:FermionScalar}}^{\mu \nu}=-\frac{T^2}{32\pi^2}\left[g_F^a g_F^b\right]^{I}_J (Y Y^c)^{J}_I\Pi_2^{\mu \nu}
-\frac{T^2}{192\pi^2}\left[g_S^a g_S^b\right]_{ij}(Y Y^c+Y^c Y)^{ij}\Pi_2^{\mu \nu}
\end{align}
\section{Scalar Diagrams}\label{app:ScalarDiagrams}
\subsection{Scalar current}
The vector sunset gives
\begin{align}
	\Pi^\mu_{\ref{fig:ScalarVector}}=\frac{1}{2}g^{a}_{ni}g^c_{jk}g^d_{lm}&\int_{PQ} F^{\mu} \left\{\delta^{kl} \delta^{mn} \Delta_S^{ij}(P)\Delta_{V}^{cd}(Q)\left[\Delta^R(P)\Delta^R(P+Q)+\Delta^A(P)\Delta^A(P+Q)\right] \right.\nonumber
	\\&\left.+\delta^{mn} \delta^{cd} \Delta_S^{ij}(P)\Delta_{S}^{kl}(P+Q)\left[\Delta^R(P)\Delta^A(Q)+\Delta^A(P)\Delta^R(Q)\right] \right.\nonumber
	\\&\left.+\delta^{ij}\delta^{nm}\Delta_{V}^{cd}(Q)\Delta^{
		kl}_{S}(P+Q)\left[\Delta^R(P)\Delta^A(P)\right] \right\},
	\\& F^\mu=4 i p^\mu \left\{(P+Q)^2+P^2-\frac{1}{2}Q^2\right\}.\nonumber
\end{align}
Since the scalar-vector bubble give the same combination of couplings we can group the diagrams together. We find
\begin{align}
\Pi^\mu_{\ref{fig:ScalarVector}}+	\Pi^\mu_{\ref{fig:ScalarVectorBubble}}=
	&-\frac{D+2}{8}\text{Tr}\left[g_S^a g_S^b g_S^c g_S^c\right]\int_{PQ}\frac{n_B(q)}{q p}\left\{\partial^p_0\left[N_S(p)-\overline{N}_S(p)\right]v^\mu_p-\frac{(v^\mu_p-n^\mu)}{p}\left(N_S(p)-\overline{N}_S(p)\right)\right\}^{b}. \nonumber
\end{align}
After performing the integrals we obtain
\begin{align}
\Pi^{\mu\nu}_{\ref{fig:ScalarVector}}+	\Pi^{\mu\nu}_{\ref{fig:ScalarVectorBubble}}=
&\frac{T^2(D+2)}{192 \pi^2}\text{Tr}\left[g_S^a g_S^b g_S^c g_S^c\right]\Pi^{\mu \nu}_2.
\end{align}

The scalar-bubble gives
\begin{align}
	\Pi^\mu_{\ref{fig:ScalarScalarBubble}}=\frac{1}{4}g^{a}_{li}\lambda^{jlmn}&\int_{PQ} F^{\mu} \delta^{kl}  \Delta_S^{ij}(P)\Delta_{S}^{mn}(Q)\left[\Delta^R(P)+\Delta^A(P)\right] \nonumber
	\\& F^\mu=2p^\mu,
\end{align}
which simplify to
\begin{align*}
	\Pi^\mu_{\ref{fig:ScalarScalarBubble}}=\frac{1}{8}\left[g_S^a g_S^b\right]_{jl}\lambda^{jlnn}\int_{PQ}\frac{n_B(q)}{q p}\left\{\partial^p_0\left[N_S(p)-\overline{N}_S(p)\right]v^\mu_p-\frac{(v^\mu_p-n^\mu)}{p}\left(N_S(p)-\overline{N}_S(p)\right)\right\}^{b}.
\end{align*}
After performing the integrals we find
\begin{align}
	\Pi^{\mu\nu}_{\ref{fig:ScalarScalarBubble}}=-\frac{T^2}{192 \pi^2}\text{Tr}\left[g_S^a g_S^b\right]_{jl}\lambda^{jlnn}  \Pi^{\mu \nu}_2.
\end{align}

We can also have scalar-mass insertions from one loop diagrams:
\begin{align}
	\Pi^\mu_{\ref{fig:ScalarMass}}=ig^{a}_{ki} \mu^{jk}\int_P \Delta_S^{ij}(P)\left(\Delta^R(P)+\Delta^R(P)\right),
\end{align}
which gives
\begin{align}
	\Pi^\mu_{\ref{fig:ScalarMass}}=\left[g_S^a g_S^b\right]_{ij}\mu^{jj}\int_P\frac{1}{p}\left\{\partial^p_0\left[N_S(p)-\overline{N}_S(p)\right]v^\mu_p-\frac{(v^\mu_p-n^\mu)}{p}\left(N_S(p)-\overline{N}_S(p)\right)\right\}^{b}.\nonumber
\end{align}
After performing the integral we find
\begin{align}
\Pi^{\mu \nu}_{\ref{fig:ScalarMass}}=-\frac{1}{8 \pi^2}\text{Tr}\left[g_S^a g_S^b\right]_{ij}\mu^{jj} \Pi_2^{\mu \nu}.
\end{align}

\subsection{Vector current}
The scalar sunset gives
\begin{align}
	\Pi^\mu_{\ref{fig:VectorScalar}}=\frac{1}{2}g_V^{ace}g^d_{jn}g^f_{mi}&\int_{PQ} F^{\mu} \left\{\delta^{nm} \delta^{ef} \Delta_V^{cd}(P)\Delta_{S}^{ij}(Q)\left[\Delta^R(P)\Delta^R(P+Q)+\Delta^A(P)\Delta^A(P+Q)\right] \right.\nonumber
	\\&\left.+\delta^{ef} \delta^{ij} \Delta_V^{cd}(P)\Delta_{S}^{nm}(P+Q)\left[\Delta^R(P)\Delta^A(Q)+\Delta^A(P)\Delta^R(Q)\right] \right.\nonumber
	\\&\left.+\delta^{ef}\delta^{cd}\Delta_{S}^{ij}(Q)\Delta^{
		nm}_{S}(P+Q)\left[\Delta^R(P)\Delta^A(P)\right] \right\},
	\\& F^\mu=i\left[(4 q^\mu-2p^\mu)(P+Q)^2+2 P^2 p^\mu -Q^2(6 p^\mu+2 q^\mu)\right],\nonumber
\end{align}
or after simplifying
\begin{align}
	&\Pi^{\mu\nu}_{\ref{fig:VectorScalar}}=- g_V^{aec}g_V^{bdc}\text{Tr}\left[g_S^d g_S^f\right] \int_{PQ}\frac{n_B(q)}{q p}\left\{\partial^p_0\left[N_S(p)-\overline{N}_S(p)\right]v^\mu_p-\frac{(v^\mu_p-n^\mu)}{p}\left(N_S(p)-\overline{N}_S(p)\right)\right\}^{b} \nonumber
	\\&+\frac{1}{4} g_V^{aec}g_V^{bdc}\text{Tr}\left[g_S^d g_S^f\right]\int_{PQ}(N_V(q)-\overline{N}_V(q))^b \frac{v^\mu_q}{p^2}\left\{d_0(N_S+\overline{N}_S)-\frac{1}{p}(N_S+\overline{N}_S)\right\}
\end{align}
After performing the integrals we find
\begin{align}
\Pi^{\mu\nu}_{\ref{fig:VectorScalar}}&=\frac{T^2}{24 \pi^2}  g_V^{aec}g_V^{bdc}\text{Tr}\left[g_S^d g_S^e\right] \Pi_2^{\mu \nu} -T^2 I_\text{SV}  g_V^{aec}g_V^{bdc}\text{Tr}\left[g_S^d g_S^e\right] \Pi_1^{\mu \nu},
\end{align}
where
\begin{align}
I_\text{SV}=\int_{PQ} n'_B(q)\frac{1}{p^2}\left(n'_B(p)-\frac{n_B(p)}{p}\right)=\left\{\frac{1}{48 \pi^2 \epsilon}+\frac{T^2 (24 \log (A)+4 \log \frac{\mu}{4\pi T}+2 \gamma -1)}{48 \pi ^2}\right\},
\end{align}

Finally, the scalar bubble gives
\begin{align}
	\Pi^\mu_{\ref{fig:VectorScalarBubble}}=\frac{1}{2}g_V^{ace}H_{V,ij}^{df}&\int_{PQ} F^{\mu} \delta^{ef}  \Delta_V^{cd}(P)\Delta_{S}^{ij}(Q)\left[\Delta^R(P)+\Delta^A(P)\right], \quad F^\mu=-2(D-1)p^\mu,\nonumber
\end{align}
which after simplifying gives
\begin{align}
	\Pi^\mu_{\ref{fig:VectorScalarBubble}}=\frac{(D-1)}{2}g _V^{aec}g_V^{bdc}\text{Tr}\left[g_S^d g_S^e\right]\int_{PQ}\frac{n_B(q)}{q p}\left\{\partial^p_0\left[N_V(p)-\overline{N}_V(p)\right]v^\mu_p-\frac{(v^\mu_p-n^\mu)}{p}\left(N_V(p)-\overline{N}_V(p)\right)\right\}^{b}\nonumber.
\end{align}
After doing the integrals we find
\begin{align}
	\Pi^{\mu\nu}_{\ref{fig:VectorScalar}}+\Pi^{\mu \nu}_{\ref{fig:VectorScalarBubble}}=-T^2 \frac{(D-3)}{48 \pi^2}   g_V^{aec}g_V^{bdc}\text{Tr}\left[g_S^d g_S^e\right] \Pi_2^{\mu \nu}-T^2 I_\text{SV}  g_V^{aec}g_V^{bdc}\text{Tr}\left[g_S^d g_S^e\right] \Pi_1^{\mu \nu}.
\end{align}

\section{Counter-term contributions}\label{app:CounterTerms}
To renormalize we need wave-function and coupling counter-terms. These are all well known~\cite{Machacek:1983tz,Machacek:1983fi,Machacek:1984zw}. The anomalous dimensions are\footnote{We here omit all Yukawa couplings since their counter-term contributions cancel.}
\begin{align}
&\gamma^I_J=\frac{1}{16\pi^2}\left\{-\left[g_F^c g_F^c\right]^{I}_J\right\},
\\& \gamma_V^{ab}=\frac{1}{16 \pi^2}\left\{-\frac{5}{3}\text{Tr}[g_V^a g_V^b]-\frac{2}{3}\text{Tr}[g_F^a g_F^b]+\frac{1}{6}\text{Tr}\left[g^a_S g^b_S\right]\right\},
\\&  \gamma_g^{ab}=-\frac{1}{16 \pi^2}\left\{ \frac{1}{2}\text{Tr}\left[g_V^a g_V^b\right]\right\}, \quad \gamma_S^{ij}=\frac{1}{16\pi^2}\left\{-2\left[g^c g^c\right]^{ij}\right\}.
\end{align}

Next the vector and fermion-vector trilinear couplings:
\begin{align}
&\delta g^{a,I}_J=\frac{1}{32 \pi^2 \epsilon}\left\{-2\left[g_F^c g_F^a g_F^c\right]^I_J+6 i g_V^{abc} \left[g_F^b g_F^c\right]^I_J  - \gamma_V^{ab} g_{F,J}^{b,I}-g_{F,K}^{a,I} \gamma^{K}_J-g_{F,J}^{a,K} \gamma^{*,I}_K\right\}.
\\& \delta g^{abc}=\frac{1}{32 \pi^2 \epsilon}\left\{-2 \text{Tr}\left[g^a_V g^b_V g^c_V\right] -g_V^{abe}\gamma_V^{ec}-\gamma_g^{ae}g_V^{ebc}-\gamma_g^{be}g_V^{aec} \right\}.
\end{align}
For the scalar-coupling we only need the combination $H^{ab}_{ij}=g^a_{ik}g^b_{kj}+g^b_{ik}g^a_{kj}$. The counter-term is
\begin{align}
\delta H^{ab}_{ij}=-\frac{1}{16\pi^2 \epsilon} &\left\{\frac{8}{3} g_V^{ace}g_V^{bef}H^{cf}_{ij}+2\left[g_S^c(g_S^a g_S^b+g_S^b g_S^a) g_S^c\right]_{ij} \right.
\\&\left. -\frac{1}{2}\left[\gamma_S^{in}H^{ab}_{nj}+\gamma_S^{jn}H^{ab}_{in}+\gamma_V^{ac}H^{cb}_{ij}+\gamma_V^{bc}H^{ac}_{ij}\right]  \right\}
\end{align}

\subsection{Vector loops}
Using the counter-terms from section \ref{app:CounterTerms} we find
\begin{align}
\Pi_\text{CT,V}^{\mu \nu}=-\frac{(D-2)}{4\pi^2 \epsilon}g^{acd}g^{def}g^{cfn}g^{bnx} \int_P n_B'(p) \Pi_1^{\mu \nu},
\end{align}
where
\begin{align}
\int_P n_B'(p)=-\frac{T^2}{6}-\frac{T^2}{6}  \left(24 \log (A)+2 \log \frac{\mu}{4\pi T}-1\right)\epsilon+\mathcal{O}(\epsilon^2).
\end{align}

\subsection{Fermion loops}
Using the counter-terms from section \ref{app:CounterTerms} we find
\begin{align}
	\Pi_\text{CT,F}^{\mu \nu}=-\frac{1}{4\pi^2 \epsilon}g_V^{ace}g_V^{bcd}\text{Tr}\left[g_F^d g_F^e\right] \int_P n_F'(p) \Pi_1^{\mu \nu},
\end{align}
where
\begin{align}
	\int_P n_F'(p)=-\frac{T^2}{12}-\frac{T^2}{12}  \left(24 \log (A)+2 \log \frac{\mu}{4\pi T}-1-\log 4\right)\epsilon+\mathcal{O}(\epsilon^2).
\end{align}
\subsection{Scalar loops}
Using the counter-terms from appendix \ref{app:CounterTerms} we find
\begin{align}
	\Pi_\text{CT,S}^{\mu \nu}=\frac{1}{8\pi^2 \epsilon}g _V^{aec}g_V^{bdc}\text{Tr}\left[g_S^d g_S^e\right] \int_P n_B'(p) \Pi_1^{\mu \nu}.
\end{align}

\subsection{Power corrections before field redefinitions}\label{app:PowerFull}
The scalar loop gives
\begin{align}
	&g_{\mu \nu }\Pi^{\mu \nu,ab}_{S}(K)=\text{Tr}\left[g_S^a g_S^b\right]\frac{K^2}{16 \pi^2}\left\{\frac{1}{2\epsilon}+\log\frac{\mu e^\gamma}{4\pi T}+k^0 L(K)\right\},
	\\& \Pi^{00,ab}_S(K)=-\text{Tr}\left[g_S^a g_S^b\right]\frac{1}{3}\frac{k^2}{16 \pi^2}\left\{\frac{1}{2 \epsilon}+\log\frac{\mu e^\gamma}{4\pi T}+1+ \frac{(k^0)^2}{k^2} (k^0 L(K)-1)\right\}
\end{align}

The fermion loop gives
\begin{align}
	&g_{\mu \nu }\Pi^{\mu \nu,ab}_{F}(K)=\text{Tr}\left[g_F^a g_F^b\right]\frac{K^2}{16 \pi^2}\left\{\frac{2}{\epsilon}+4 \left(\log\frac{\mu e^\gamma}{4\pi T}+\log 4\right)-2+4 k^0 L(K)\right\},
	\\& \Pi^{00,ab}_F(K)=-\text{Tr}\left[g_F^a g_F^b\right]\frac{1}{3}\frac{k^2}{16 \pi^2}\left\{\frac{2}{\epsilon}+4 \left(\log\frac{\mu e^\gamma}{4\pi T}+\log4\right)-2+2k^0 \left(3- \frac{(k^0)^2}{k^2}\right)L(K)+2\frac{(k^0)^2}{k^2}\right\}.\nonumber
\end{align}

For non-abelian diagrams we group ghosts and vectors together, the result is
\begin{align}
	&g_{\mu \nu }\Pi^{\mu \nu,ab}_{V}(K)=\text{Tr}\left[g_V^a g_V^b\right]\frac{K^2}{16 \pi^2}\left\{\frac{5}{\epsilon}+10\log\frac{\mu e^\gamma}{4\pi T}-3+10 k^0 L(K)\right\},
	\\& \Pi^{00,ab}_V(K)=-\text{Tr}\left[g_V^a g_V^b\right]\frac{1}{3}\frac{k^2}{16 \pi^2}\left\{\frac{5}{\epsilon}+10\log\frac{\mu e^\gamma}{4\pi T}-1+2 k^0 \left(6-\frac{(k^0)^2}{k^2}\right)L(K)+2\frac{ (k^0)^2}{ k^2}\right\}.\nonumber
\end{align}

\bibliographystyle{utphys}

{\linespread{0.6}\selectfont\bibliography{Bibliography}}

\end{document}